\documentclass{aa}
\usepackage{graphicx}
\usepackage{epsfig,graphicx,rotating,portland,fancyheadings,longtable,natbib}
\usepackage{multirow}
\usepackage{txfonts}

\def\gs{{_>\atop^{\sim}}}

\begin{document}
   \title{UVES/VLT high resolution absorption spectroscopy of the
GRB080330 afterglow: a study of the GRB host galaxy and intervening
absorbers \thanks{Based on observations collected at the European
Southern Observatory, ESO, the VLT/Kueyen telescope, Paranal, Chile,
in the framework of programs 080.A-0398}}

\author{V. D'Elia$^1$, F. Fiore$^1$, R. Perna$^{2}$, Y. Krongold$^{3}$, S.D. Vergani$^{4}$, S. Campana$^5$,
S. Covino$^5$, P. D'Avanzo$^5$, D. Fugazza$^5$, P. Goldoni$^{6,7}$, C. Guidorzi$^5$, E.J.A. Meurs$^{8,9}$, 
L. Norci$^9$, S. Piranomonte$^1$, G. Tagliaferri$^5$, P. Ward$^{10}$
}

\institute
{$^1$ INAF-Osservatorio Astronomico di Roma, Via Frascati 33, I-00044 Monteporzio Catone, Italy;\\ 
$^2$ JILA, Campus Box 440, University of Colorado, Boulder, CO 80309-0440, USA; \\
$^3$ Instituto de Astronomia, Universidad Nacional Autonomica de 
Mexico, Apartado Postal 70-264, 04510 Mexico DF; \\
$^4$ University Paris 7, APC, Lab. Astroparticule et Cosmologie, UMR7164 CNRS, 10 rue Alice 
Domon et LÂonie Duquet 75205 Paris Cedex 13;  \\
$^5$ INAF, Osservatorio Astronomico di Brera, via E. Bianchi 46, 23807 Merate (LC), Italy;\\ 
$^6$ Laboratoire Astroparticule et Cosmologie, 10 rue A. Domon et L. Duquet, 75205 Paris Cedex 13, France;\\
$^7$ Service d'Astrophysique, DSM/DAPNIA/SAp, CEA-Saclay, 91191 Gif-sur-Yvette, France;\\
$^8$ School of Cosmic Physics, DIAS, 31 Fitzwilliam Place, Dublin 4, Ireland;\\
$^9$ School of Physical Sciences and NCPST, DCU, Glasnevin, Dublin 9, Ireland;\\
$^{10}$ Dunsink Observatory, Castleknock, Dublin 15, Ireland;\\ 
}

  \abstract 
  {} 
  {We study the Gamma Ray Burst (GRB) 
  environment and intervening absorbers by
  analyzing the optical absorption features produced by gas
  surrounding the GRB or along its line of sight.  
  }  
  {We analyzed high
  resolution spectroscopic observations (R=40000, S/N$=3 - 6$) of
  the optical afterglow of GRB080330, taken with UVES at the VLT $\sim 1.5$
  hours after the GRB trigger.}
  {The spectrum illustrates the complexity of the ISM of the GRB host
  galaxy at z = 1.51 which has at least four components in
  the main absorption system. We detect strong {\ion{Fe}{II}},
  {\ion{Si}{II}}, and {\ion{Ni}{II}} excited absorption lines
  associated with the bluemost component only.  In addition to the
  host galaxy, at least two more absorbers lying along the line of
  sight to the afterglow have been detected in the redshift range $
  0.8 < z < 1.1$, each exhibiting {\ion{Mg}{II}} absorption. For the
  bluemost component in the host galaxy, we derive information about its
  distance from the site of the GRB explosion.  We do so by assuming
  that the excited absorption lines are produced by indirect UV
  pumping, and compare the data with a time dependent photo-excitation
  code.  The distance of this component is found to be $ \sim
  280^{+40}_{-50}$ pc, which is lower than found for other GRBs
  ($ 1 - 6 $ kpc).  We identify two additional {\ion{Mg}{II}}
  absorbers, one of them with a rest frame equivalent width larger
  than 1\AA. }
  {The distance between the GRB and the absorber measured in this paper
  confirms that the power of the GRB radiation can influence the
  conditions of the interstellar medium up to a distance of at least
  several hundred pc.
  
  For the intervening absorbers, we confirm the trend
  that on average one strong intervening system is found per
  afterglow, as has been noted in studies exhibiting an excess of strong
  {\ion{Mg}{II}} absorbers along GRB sightlines compared to
  quasars.  }

\authorrunning {D'Elia et al.}
\titlerunning {UVES/VLT high resolution absorption spectroscopy of GRB080330}

\maketitle
%

\section{Introduction}

The study of the interstellar medium (ISM) of z$\gs1$ galaxies has so
far relied upon observation of Lyman-break galaxies (LBGs) at z$=3-4$
(see e.g. Steidel et al.  1999) and galaxies that happen to be along
lines of sight to bright background quasars (or QSOs, see e.g.,
Prochaska et al. 2003).  However, LBGs are characterized by
significant star formation and their inferred chemical abundances may
not be representative of typical high-z galaxies.  Weak metal-line
systems along the line of sight to quasars probe mainly galaxy haloes,
rather than their bulges or discs. Taking advantage of ultra-deep {\it
Gemini} multi-object spectrograph observations, Savaglio et al. (2004,
2005) studied the ISM of a sample of faint K-band selected galaxies at
1.4$<$z$<$2.0, finding MgII and FeII abundances much higher than in
QSO systems but similar to those in GRB host galaxies (Prochaska et
al. 2007). These studies can hardly be extended to higher redshift
with the present generation of 8m class telescopes, because of the
faintness of high-z galaxies. Since the discovery that Gamma-Ray
Bursts (GRBs) are extragalactic, we are now able to use an independent
tool to study the ISM of high-z galaxies.


The main difference between QSO and GRB absorption spectroscopy is
that QSOs are stationary in their emission, while GRBs are the most
variable and violent phenomena in the Universe known to date. Thus,
while the QSO has already ionized the ISM along its line of sight, the
physical, dynamical, and chemical status of the circumburst medium in
the star-forming region hosting the GRB progenitor can be modified by
the explosive event, by shocks and ionizing photons. The
presence of blue-shifted absorbers may indicate that the GRB and its
afterglow can not only ionize the interstellar medium, but further
radiatively and/or collisionally accelerate it. Circumstellar
absorption lines in GRB afterglow spectra can be used to determine the
main properties of the stellar progenitors (van Marle, Langer, \&
Garcia-Segura 2005). The variability in the intensity of absorption
lines can be used to determine the location and density of the
absorber, following Perna \& Loeb (1998), Boettcher, Fryer \& Dermer
(2002), Draine \& Hao (2002), Mirabal et al. (2002), Perna \& Lazzati
(2002), and Perna, Lazzati \& Fiore (2003).

Occasionally, extremely bright optical transient
emission is associated with a GRB event, offering a superb
opportunity to investigate high--z galaxies by means of high resolution
spectroscopy of the optical transient.

Fiore et al. (2005) presented the first high-resolution spectroscopic
data of GRB afterglows. Their study of the UVES/VLT spectra of
GRB020813 and GRB021004 showed that the ISM of GRB host galaxies is
complex when resolved down to a width of a few tens of km/s, many
velocity components contributing to each main absorption system and
spanning a total velocity range of up to thousands of km/s.


Since the launch of {\it Swift}, which now allows a prompt reaction to
a GRB event, a dozen afterglows have been observed at high
resolution. The first published high resolution observations in the
{\it Swift} era were those of the GRB051111 and GRB050730 afterglows
(Chen et al. 2005; Prochaska, Chen \& Bloom 2006; Penprase et
al. 2006).  They confirm the complex nature of the ISM of the GRB host
galaxies, and velocity ranges of hundreds of km/s are found with the
spectra obtained with MIKE at the MagellanII and HIRES at the
Keck. GRB050730 was also observed using UVES at the VLT (D'Elia et
al. 2007): the ISM is inferred to have a clumpy nature from the
presence of six components in the high resolution spectrum, each
representing a different layer of absorbing gas. The capability of
resolving the absorbing gas into components of different velocity
allows us to separate the contribution of the circumburst environment
from that of regions in the host galaxy far away and less affected by
the GRB afterglow, as in the case of GRB 050922C (Piranomonte et
al. 2008). Fine structure and other excited levels of the species
{\ion{O}{I}} and the ions {\ion{Fe}{II}}, {\ion{Ni}{II}},
{\ion{Si}{II}}, and {\ion{C}{II}}, are routinely identified in GRB
spectra. These lines are most probably excited by the intense UV flux
coming from the afterglow, since strong variation is observed when
multi-epoch, high resolution spectroscopy is available.  This
variation is inconsistent with a pure infrared excitation or
collisional processes (Vreeswijk et al. 2007, D'Elia et
al. 2009). Thus, by assuming UV pumping as the responsible mechanism
for the production of these lines, we can compute the distance of the
gas from the GRB. This distance is found to be of the order of a few
kpc (see Vreeswijk et al. 2007 for GRB060418) or even larger, as in
the case of the naked-eye GRB080319B (D'Elia et al. 2009), for which
the inferred distance of 6 kpc to the farthermost component allows
speculation about a possible extragalactic origin of part of the
excited gas.

GRB high resolution spectroscopy is also suitable for studying the
systems lying along the line of sight to GRBs, and not only their host
galaxies. Surprisingly, the number density of {\ion{Mg}{II}}
intervening absorbers in GRB spectra is nearly 4 times higher than
that along QSO sightlines (Prochter et al. 2006), while the
{\ion{C}{IV}} absorbers do not show any statistical difference
(Sudilovsky et al. 2007; Tejos et al. 2007). The reason for the
{\ion{Mg}{II} excess in GRB spectra} is still uncertain, and a larger
sample is required to properly address this issue.

In this paper, we discuss high resolution spectroscopy of GRB080330.
We investigate both the local medium surrounding the GRB, and the
intervening systems. For what concerns the host galaxy, we focus
in particular on the analysis of the absorption lines produced by the
excited levels of the atoms and ions constituting the absorbing
gas. The comparison of these absorption features and their
corresponding ground states yields information about the gas distance
from the GRB explosion site, based on the assumption of indirect UV pumping
production of the excited levels. A full analysis of the intervening
absorbers lying along the line of sight to the GRB afterglow is also
presented.

The paper is organized as follows. Section $2$ summarizes
of the GRB080330 detection and observations. Section $3$ presents the
high resolution observations and data reduction. Section $4$ is
devoted to the study of the absorbers located in the host galaxy, and
in particular to the analysis of their excited levels. Section $5$
presents a full analysis of the absorbing systems identified in the
spectrum. Finally, in Sect. $6$ the results are discussed and
conclusions are drawn.

\section{GRB080330}

GRB080330 was discovered by {\it Swift}-BAT on March 30, 2008, at
03:41:16 UT (Mao et al. 2008). Less than 100s later, both XRT and UVOT
detected the afterglow, which had a magnitude of 18.8 (white
filter). Even before the distribution of the XRT position, the TAROT
and ROTSE robotic telescopes had detected the optical afterglow at the
position RA: 11:17:04.5 Dec: +30:37:24.0 (J2000). The magnitude of the
object was roughly 17.5 22s after the trigger, and reached a peak
of R=16.8 at 300s (Klotz et al. 2008; Schaefer \& Guver 2008). The
redshift of the host galaxy was first secured by NOT observations
(Malesani et al. 2008), and then confirmed by Cucchiara et al. (2008),
using Hobby-Eberly Telescope observations. The afterglow was also
detected in the infrared band (J, H, and K) by PAIRITEL (Bloom \& Starr
2008) and GROND (Clemens et al. 2008), and in the $\gamma$-band (20
keV - 1 MeV) by Konus-Wind (Golenetskii et al. 2008). For a detailed
analysis of the multi-band properties of the prompt and afterglow
emission evolution, we refer the reader to Guidorzi et al (2009).

\section{Observations and data reduction}

In the framework of the ESO program 080.A-0398, we observed the
afterglow of GRB080330 with the high resolution UV-visual echelle
spectrograph (UVES, Dekker et al. 2000), mounted at the VLT-UT2
telescope. Observations began on 30 March 30 2008 at 05:06:32 UT
($\sim 1.5$h after the GRB) for a net exposure time of 45 minutes,
when the magnitude of the afterglow was $R\sim18.8$. Because of 
observational and scheduling constraints, only the UVES dichroic 1 was
used, when employing the red as well as the blue arm. The slit width was
set to be 1'' (which corresponds to a resolution of R=40000) and the
read-out mode was rebinned to 2 $\times$ 2 pixels. The achieved
spectral range covers the bands from $\sim$3300\AA\ to $\sim$3870\AA,
$\sim$4780\AA\ to $\sim$5750\AA, and $\sim$5830\AA\ to
$\sim$6810\AA. The data below $\sim$3500\AA\ is dominated totally by
noise. Table 1 provides a summary of the UVES setups used.

\begin{table*}[ht]
\caption{\bf UVES setups}
{\footnotesize
\smallskip
\begin{tabular}{|l|c|c|c|c|}
\hline 
Setup (nm)       & Wavelength (\AA)&  Slit width & Spec. Resolution &  S/N        \\
\hline 
Dic 1, 346       & 3300 - 3870     &  1''        & 40 000            &  $\sim3$    \\
\hline
Dic 1, 580       & 4780 - 6810     &  1''        & 40 000            &  $\sim6$    \\
\hline
\end{tabular}
}
\end{table*}

The data reduction was performed using the UVES pipeline
(Ballester et al. 2000). The final resolution was $10.5$ km/s for the
blue arm and $7.5$ km/s for the red arm; this slight loss of resolution in
the blue band is caused by a rebinning to increase the signal-to-noise
ratio per pixel, which ranges from $\sim 3$ at 3800\AA\ to $\sim 6$
at 6800\AA. The noise spectrum, used to determine the errors in the
best-fit line parameters, was calculated from the real,
background-subtracted spectrum, using line-free regions.
This, therefore, takes into account both statistical and systematic
errors in the pipeline processing as well as the background
subtraction.

In the following two sections, we discuss the absorption features
related to the main system, i.e., the host galaxy (Sect. 4), and
those produced by other intervening systems (Sect. 5).

\section{The main absorption system}

This system has the greatest number of features and is produced by the
gas residing in the GRB host galaxy. The spectrum exhibits a large
number of metal-absorption lines, of both neutral ({\ion{Mg}{I}}) and
low ionization species ({\ion{Si}{II}}, {\ion{Zn}{II}},
{\ion{Cr}{II}}, {\ion{Ni}{II}}, {\ion{Fe}{II}}, {\ion{Mn}{II}}). In
addition, strong absorption from the fine structure levels of
{\ion{Fe}{II}} and {\ion{Si}{II}} and from the metastable levels of
the {\ion{Fe}{II}} and {\ion{Ni}{II}} has been identified, suggesting
that the intense radiation field from the GRB excites these
features. Table 2 summarizes all the absorption lines of the main
system. The ISM of the host galaxy is complex, with many components
contributing to the absorption system. The wealth of metal-line
transitions allows us to precisely determine the redshift of the GRB
host galaxy.  This yields a heliocentric redshift of z=1.5115, if we
set the reference point to the bluemost component of the main system;
this component is the only one for which there is significant
absorption from the excited levels of the intervening gas (see
subsect. 4.2). Our redshift value agrees with that found by Malesani
et al. (2008) and Cucchiara et al. (2008). This complex environment is
described in the next following sections with a detailed study of the
excited lines that aims to estimate the distance of the circumburst gas.

\subsection{Line-fitting procedure}

The analysis of the GRB environment is complicated by the complexity
of the absorption lines in the spectrum, which in several cases cannot
easily be fitted with a single line profile. This implies that many
different velocity components contribute to the gas in the GRB
environment.  In other words, several layers of gas that may be close
to or far from each other, appear to be mixed together in the spectrum (in
velocity space).  The presence of several components is thus
indicative of clumpy gas in the GRB environment, which consists of different
absorbing regions each with different physical properties.

If the absorption features are fitted with a single Voigt profile, we
are able to obtain only very poor fitting parameters and very high $\chi^2$
values.  The identification of the different components is somewhat
subjective, the true message being that the geometry and kinematics of
the ISM clouds probed by the GRB line of sight are complex.  The
{\ion{Fe}{II}} lines have the widest velocity range, so we tried to use
them to identify the different components within the circumburst
matter.  Although several {\ion{Fe}{II}} absorption features are
saturated, two of them ($\lambda$2260 and $\lambda$2249) are not and
can guide the fit in identifying the wavelengths of the components. A
four-component model represents a good fit for the {\ion{Fe}{II}
lines. Thus, the redshifts of these components as well as the Doppler
parameters were fixed to fit all the saturated {\ion{Fe}{II}}
and other ion features present in the spectrum with the same model.


Figure. 1 presents all the absorption features relative to
{\ion{Fe}{II}}, with the exception of the saturated lines.  The
transitions due to the four fine structure levels of the
{\ion{Fe}{II}} ground state are marked with asterisks.  The lines
{\ion{Fe}{II} $\lambda$2332 and $\lambda$2360} are produced by the
first excited level.  The components are labeled I (redmost) to IV.
The reference zeropoint of the velocity shifts has been placed at
z=1.5115. This value was chosen to coincide with the
redshift of the bluemost component, for two reasons. First, the narrow
profile of the excited lines makes the redshift determination more
accurate. Second, the excited levels are possibly produced by the
indirect UV-pumping mechanism (see next subsection), implying that
component IV is probably the closest component to the GRB explosion
site. Figure 2 displays all the other species observed in the spectrum,
namely: {\ion{Mg}{I}}, \ion{Zn}{II}, \ion{Cr}{II}, \ion{Si}{II},
\ion{Mn}{II}, and \ion{Ni}{II}. The transition
{\ion{Si}{II}$\lambda$1533*} is produced by the first fine structure
level of the \ion{Si}{II}, and all the observed \ion{Ni}{II} lines
belong to the first excited level.} We note that all
the excited absorption features in the spectrum, produced by fine
structure and/or metastable levels of \ion{Fe}{II}, \ion{Si}{II}, and
\ion{Ni}{II}, show a significant absorption only in component IV,
i.e., our bluemost, reference component (the {\ion{Ni}{II} metastable
level features also a faint absorption in component III).  Table 2
lists all the transitions observed in the spectrum for each
species. The column densities for all the elements and ions of the host
galaxy absorbing gas, estimated using this four component model, are
also reported. The Table shows only the column densities of component
IV, which is the only one used in the following analysis, due to the
presence of the excited transitions. We emphasize here that the
{\ion{Si}{II} $\lambda$1526} line is strongly saturated, thus the
corresponding column density in Table 2, marked with an 'S', should be
interpreted with care.

As described in the introduction, the observation of excited
states in GRB absorption spectra is a quite common feature.  The next
sub-section is thus devoted to these features and to the information
that can be extracted from their analysis.


\begin{figure}
\centering
\includegraphics[angle=-90,width=9cm]{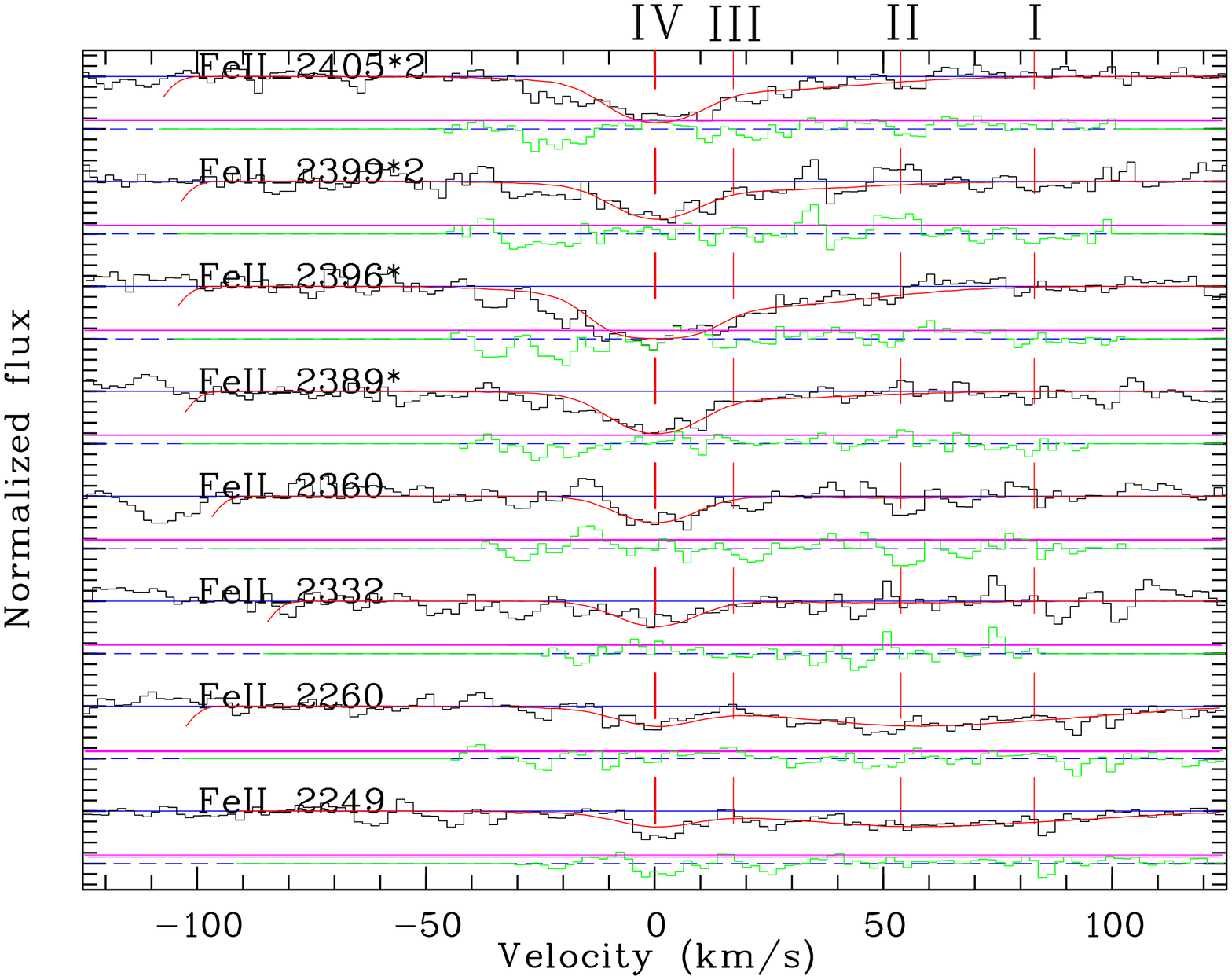}
\includegraphics[angle=-90,width=9cm]{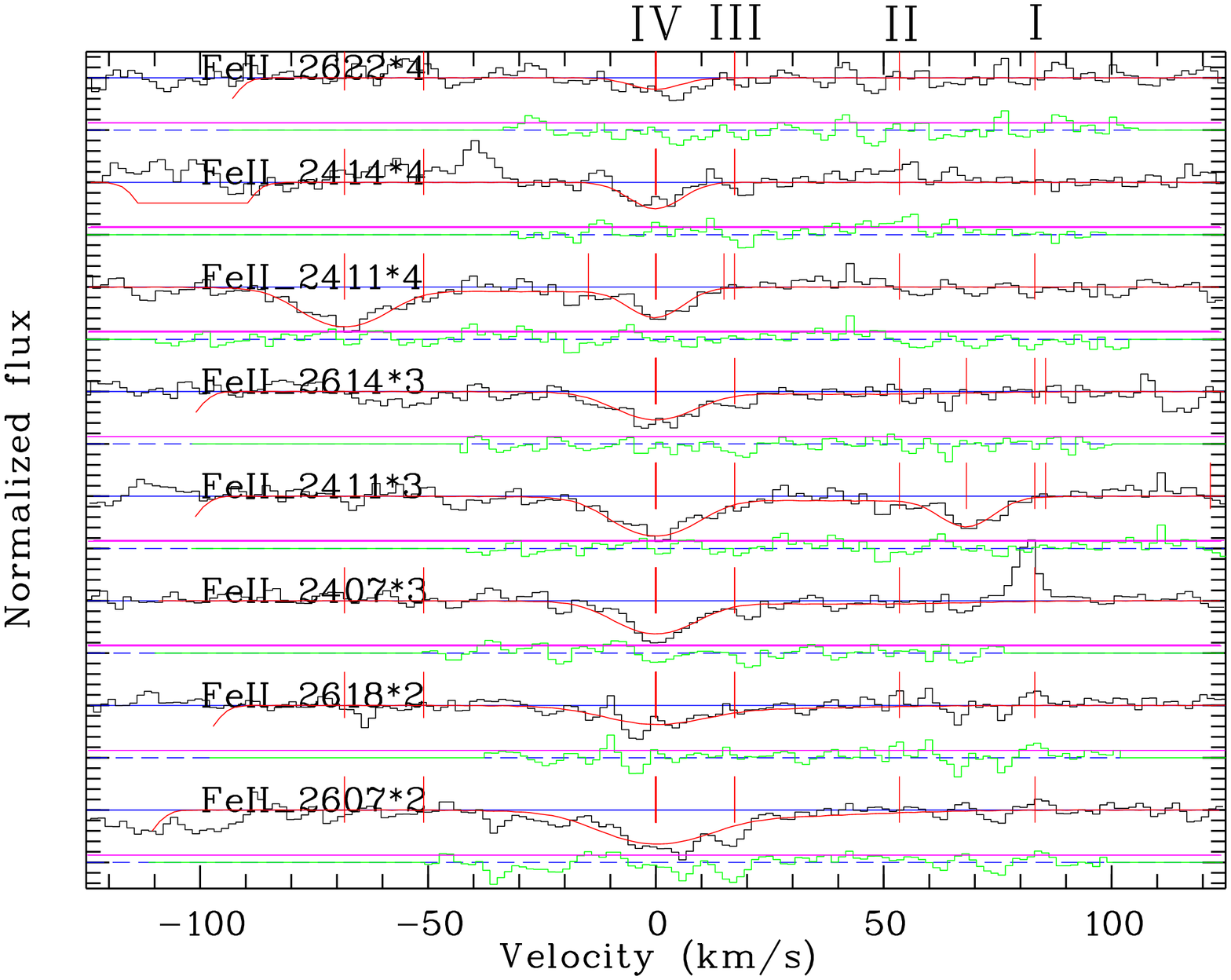}
\caption{The \ion{Fe}{II} ground and excited absorption
features. Solid lines represent our four Voigt components, best-fit
model. Vertical lines identify the velocity of each component with
respect to the zeropoint arbitrarily placed at the redshift of the 
bluemost component (z=1.5115). Bottom, solid lines mark the fit residuals.
}
\label{spe1}
\end{figure}

\begin{figure}
\centering
\includegraphics[angle=-90,width=9cm]{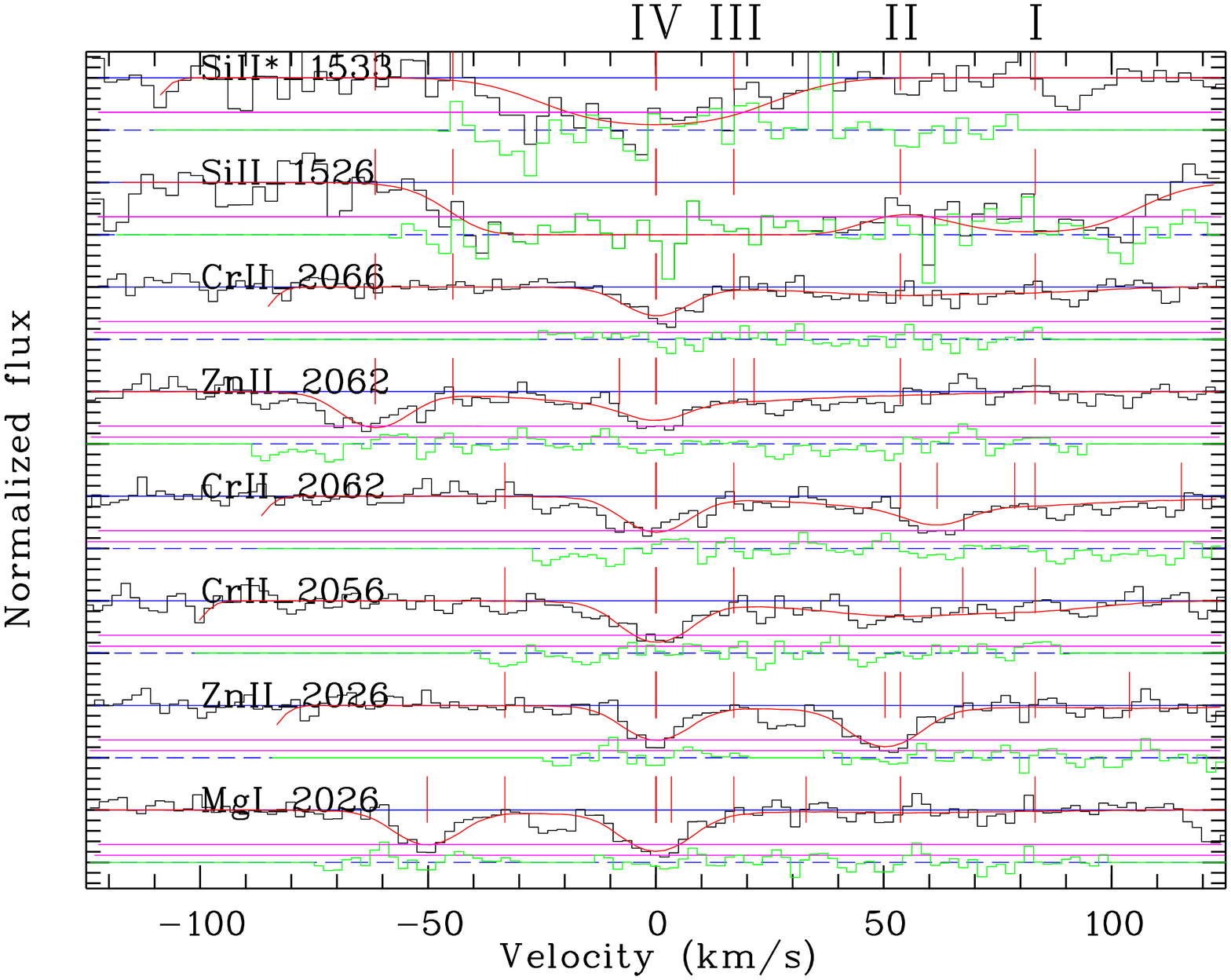}
\includegraphics[angle=-90,width=9cm]{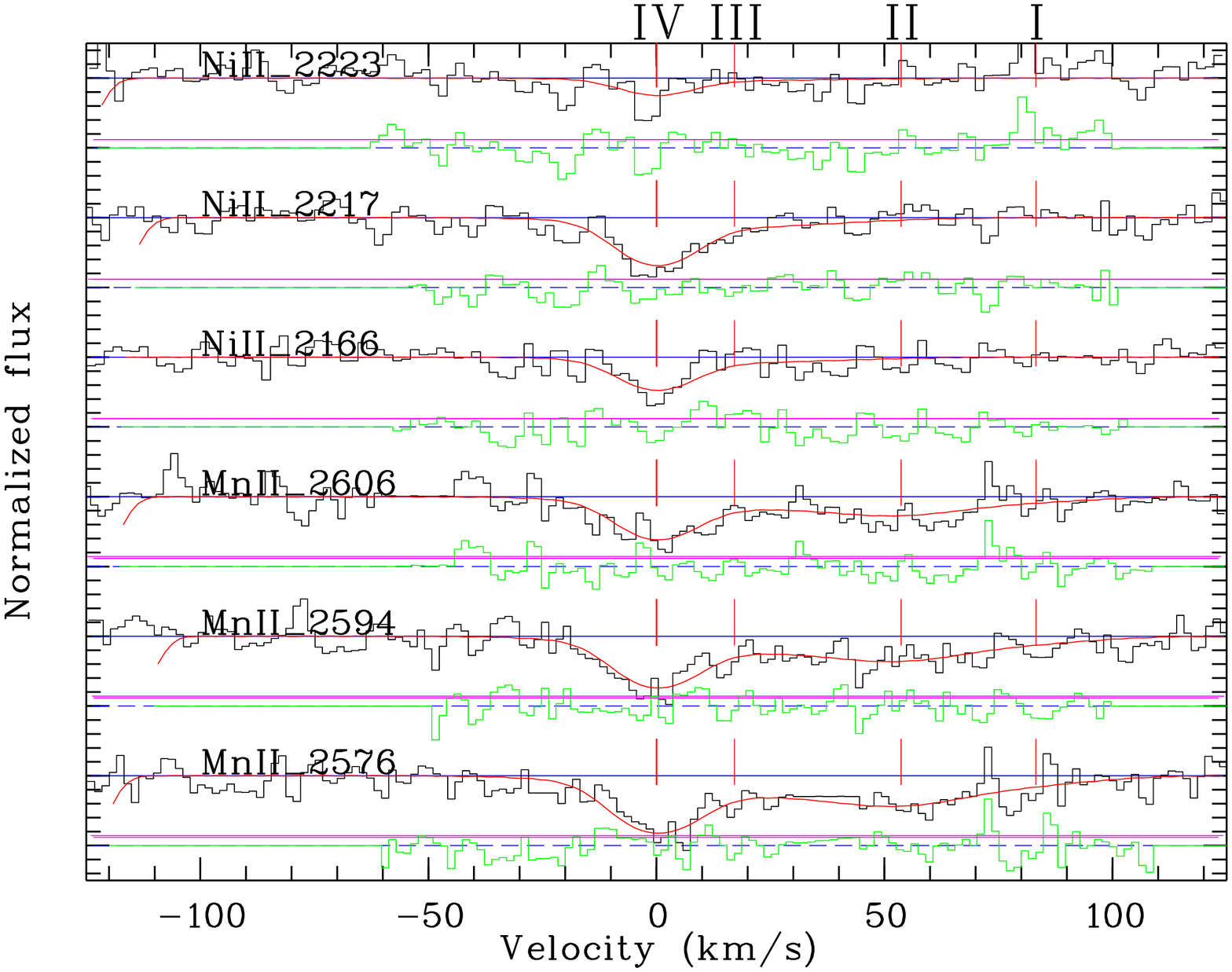}
\caption{The \ion{Mg}{I}, \ion{Zn}{II}, \ion{Cr}{II}, \ion{Si}{II}
  (top panel), \ion{Mn}{II}, and \ion{Ni}{II} (bottom panel) absorption
  features. Solid lines represent our best-fit model. Vertical lines
  identify the velocity of each component with respect to the zero
  point arbitrarily placed at the redshift of the bluemost component
  (z=1.5115). Bottom, solid lines mark the fit residuals. 
}
\label{spe1}
\end{figure}

\subsection{Excited levels}

The level structure of an atom or ion is characterized by a principal
quantum number $n$, which defines the atomic levels, and by the
spin-orbit coupling (described by the quantum number $j$), which
divides these levels into fine structrure sublevels. In GRB absorption
spectra, several excited features are detected at the GRB redshift,
because of the population of both $n>1$ and/or $n=1$ fine structure
levels. As mentioned before, component IV of the main system in the
spectrum of GRB080330 shows absorption from the four fine structure
levels of the {\ion{Fe}{II}} ground state ($a^6D$) and from the first
sublevel of the $n=2$ quantum state ($a^4F_{9/2}$). Absorption from
the {\ion{Si}{II}} $^2P^{0}_{3/2}$ fine structure level and from the
{\ion{Ni}{II}} $^4F_{9/2}$ metastable level are also present (see
Table 2 for details).
 
There are basically two mechanisms for exciting the gas of the GRB
host galaxy to these states.  The first is by collisional effects (if
the density is sufficiently high, i.e., $\ge 10^5$ cm$^{-3}$), and the
second is by the absorption of electromagnetic radiation. In the
latter case, the absorbed radiation can be from the infrared band, by
a direct population of the fine structure levels of the ground state
and other excited levels with a low value of $n$, or from the UV band,
by the population of higher levels followed by the depopulation into
the states responsible for the absorption features. Multi-epoch high
resolution spectroscopy plus the accurate modeling of the atomic level
population has proven to be a powerful tool for discriminating between
these two processes. The strong variability in the column density of
the {\ion{Fe}{II}} and {\ion{Ni}{II}} excited levels observed in
GRB060418 (Vreeswijk et al. 2007) and GRB080319B (D'Elia et al. 2009)
excluded collisional processes and direct infrared pumping as being
responsible for the excitation. Since we do not have multi-epoch
spectroscopic data, we cannot repeat the same procedure for this GRB,
but must assume that UV pumping is at work.

By assuming indirect UV pumping as the dominant mechanism for
producing the excited levels also in component IV of GRB080330, we can
compare our observed column densities to those predicted by a
photoexcitation code for the time when the high resolution observations
were acquired. This can be completed for the {\ion{Fe}{II}} and
{\ion{Si}{II}} ions, for which both ground states and excited levels
are present in the spectrum. In contrast, for {\ion{Ni}{II}},
we only measured the column density of the first metastable level,
since the transition corresponding to the ground level fell in the gap
between the blue and the red regions of the spectrum. The
photoexcitation code is that used by D'Elia et al. (2009). Basically,
it solves the detailed balance equation 

$$ {dN_u\over dt} = N_l B_{lu} F_{\nu}(\tau_0) - N_u[A_{ul} + B_{ul}F_{\nu}(\tau_0)], \eqno (1)$$

\noindent in a time-dependent way for all the transitions involving the 16
lowest levels plus 22 higher excited states (for {\ion{Fe}{II}}) and
the 20 lowest levels (for {\ion{Si}{II}}). The emission terms in that
equation depend on the flux level experienced by the absorbing
gas. This flux is of course a function of the distance of the gas from
the GRB explosion site, which is a free parameter of the
computation. The code takes into account radiative transfer and
possible opacities of some of the transitions. The other free
parameters are the initial column density of the ion, which is assumed
to be in the ground state before the GRB flux reaches it, and the
Doppler parameter of the gas. The flux of the GRB as a function of
time was measured by Guidorzi et al. (2009). Following the
notation

$$ F_R(t) = {F_R(t_0) (t/t_0)^{-\alpha} (\nu / \nu_R)^{-\beta}  \over (1+z)},  \eqno(2)$$ 

\noindent $F_R(t)$ is the flux in the R band at the time t, and the
normalization of the flux is $F_R(t_0)=4.64 \times 10^{-27}$ erg
cm$^{-2}$ s$^{-1}$ at the time $t_0=317$s. The temporal index is
$\alpha=-0.56$ before $t_0$, $\alpha=0.15$ between $t_0$ and
$t_1=1456$s and $\alpha=1.08$ for $t>t_1$, $\nu_R$ is the central
frequency of the R band. We assume here a spectral index of $\beta =
0.8$ (Guidorzi et al. 2009); this value is close to the value of
$\beta = 0.7$ inferred using the multispectral data from Cobb (2008).
The flux experienced by the absorber is

$$F^{ABS}_{\nu}(t) = (d_{L,GRB}/d)^2 \times F_{\nu}(t), \eqno (3)$$

\noindent where the GRB luminosity distance $d_{L,GRB}$ was computed
assuming an $H_0=70$ km s$^{-1}$ Mpc$^{-1}$, $\Omega_M=0.3$,
$\Omega_{\Lambda}=0.7$ cosmology. $F^{ABS}_{\nu}(t)$ was corrected by
the radiative transfer theory, before using it in eq. (1). The initial
column densities of the ground states of {\ion{Fe}{II}} and
{\ion{Si}{II}} were computed from the observed column densities of all
the levels of each ion. The exact values are: $N_{FeII} = 14.88 \pm
0.15$ and $N_{SiII} = 15.44 \pm 0.25$.  Finally, the Doppler parameter
was allowed to vary between $10$ and $20$ km s$^{-1}$, i.e., the
values that reproduce most closely the excited lines alone and all the
absorption features together, respectively. Of course all these values
refer solely to component IV, the only one for which the excited
levels are present in the spectrum. All the details of this
time-dependent photo-excitation code can be found in D'Elia et
al. (2009).

We first modeled the {\ion{Fe}{II}} photo-excitation.  The value of
the Doppler parameter that provides the closest agreement is $b=10$ km
s$^{-1}$. Figure 3 shows the model that reproduces the {\ion{Fe}{II}}
data most closely. The x-axis is the rest-frame time interval between
the GRB explosion and the UVES observation. The time of observation
(with respect to the GRB detection) is defined as the logarithmic mean
between the beginning and the end of the observation. The agreement
between the theoretical curves and the data is very good. The distance
of component IV from the GRB explosion site is measured to be $d=280$
pc.  To estimate the error associated to $d$, we ran our code for a
grid of distances in the range $50 - 800$ pc.  We then converted the
column densities and the corresponding errors to linear values to
compute the $\tilde\chi^2$.  Figure 4 plots this quantity as a function
of $d$, which has a minimum at $d=280$ pc ($\tilde\chi_{280}^2 =
0.97$). The 90\% confidence interval is enclosed in the dashed lines
and covers the range $230 < d < 320$ pc.

We fixed the model parameters (with the exception of the distance) to
those evaluated for {\ion{Fe}{II}} and ran the photoexcitation code
for {\ion{Si}{II}}. As noted in the previous subsection, the
{\ion{Si}{II}} ground state column density for component IV (Table 2)
should be interpreted with caution, since it is measured from a single
absorption line, that is saturated. In a steady-state approximation,
we would expect the ratio of the column density of {\ion{Si}{II}} fine
structure to ground levels to be $\sim 10$ times higher than the
corresponding {\ion{Fe}{II}} ratio (see e.g., Prochaska, Chen \& Bloom
2006). Thus, the {\ion{Si}{II} $\lambda$1526} value that we found is
unusual, since the ratio is $0.1$ for {\ion{Fe}{II}} while the one for
{\ion{Si}{II}} is only $0.07$. This means that we are overestimating
the {\ion{Si}{II}} ground state column density and we should consider
our value as an upper limit to the real density. The result of the
photoexcitation code for the {\ion{Si}{II}} levels is shown in
Fig. 5. The distance that we find is $d=800$ pc, which accordingly
should be considered as an upper limit consistent with the estimation
for the {\ion{Fe}{II}}. We also note that for a steady-state
approximation our {\ion{Si}{II}} column densities would imply a
distance from the GRB $3 - 5$ times higher than that inferred from
{\ion{Fe}{II}}. This is again consistent with the results of our
time-dependent code.

To ensure self-consistency, we need to establish that, at the distance
of $280$ pc as derived for component IV, \ion{Fe}{II} and
{\ion{Mg}{I}} are not fully photoionized by the strong UV radiation of
the burst. For this purpose, we performed a series of runs of the
photoionization code by Perna \& Lazzati (2002), which accounts for
the radiative-transfer of the radiation. We first simulated a medium
in thermal equilibrium at a temperature of $\sim 10^4$ K, and allowed
the radiation from the burst, modeled by Eq.(1), to propagate through
the medium for $40$ minutes in the rest frame. We found that for
absorber densities in the range of $1 - 10^3$ cm$^{-3}$, the
\ion{Fe}{II} column density is substantially unchanged after $40$
minutes. The column density of \ion{Mg}{I} is reduced by $\sim 7\%$
for densities of $\sim 1$ cm$^{-3}$, by $\sim 3\%$ for $\sim 10^2$
cm$^{-3}$, and is unaltered for $\sim 10^3$ cm$^{-3}$. We can thus
safely conclude that there is a wide range of densities for which
\ion{Fe}{II} and \ion{Mg}{I} are not substantially photoionized if the
GRB is at $280$ pc.

\begin{figure}
\centering
\includegraphics[width=9cm, height=9cm, angle=-0]{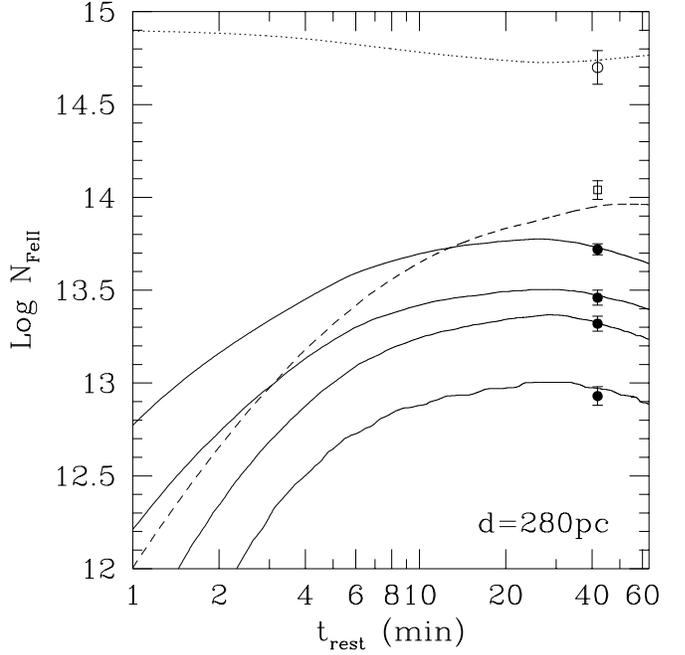}
\caption{\ion{Fe}{II} column densities for the ground level (open
circle), fine structure level (solid circles), and first excited level
(square) transitions for component IV in the spectrum of
GRB080330. Column density predictions from our time-dependent
photo-excitation code are also shown. They refer to the ground level
(dotted line), fine structure levels (solid lines), and first excited
level (dashed line) transitions, in the case of an absorber placed at 280 pc
from the GRB.
}
\label{spe1}
\end{figure}

\begin{figure}
\centering
\includegraphics[width=9cm, height=9cm, angle=-0]{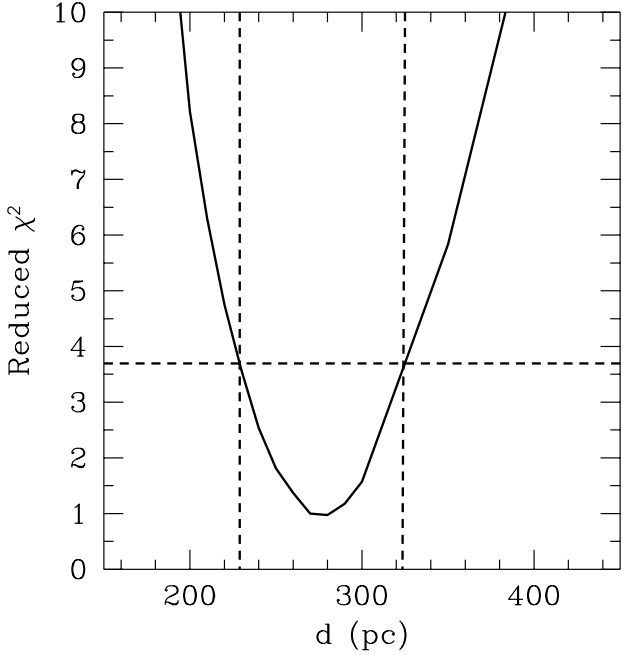}
\caption{The $\tilde\chi^2$ as a function of the distance of the
absorber from the GRB. It defines the quality of the agreement between
the observed \ion{Fe}{II} column densities and that predicted by our
time-dependent photo-excitation code (see also fig. 3). The
$\tilde\chi^2$ has a minimum at $d=280$ pc, and the 90\% confidence
distance range is enclosed by dashed lines.
}
\label{spe1}
\end{figure}

\begin{figure}
\centering
\includegraphics[width=9cm, height=9cm, angle=-0]{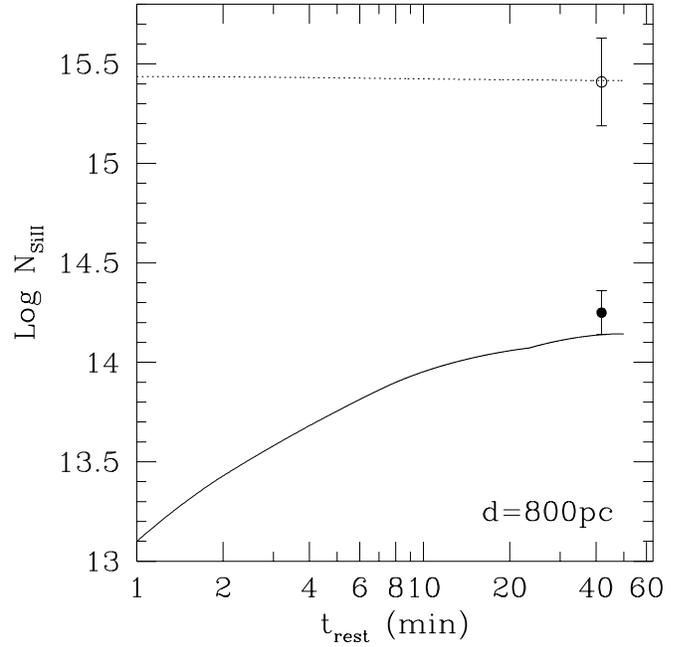}
\caption{The column densities for the ground level of \ion{Si}{II}
(open circle) and fine structure transitions for component IV in the
spectrum of GRB080330. Column density predictions from our
time-dependent photo-excitation code are also shown. They refer to the
ground level (dotted line) and fine structure level (solid line)
transitions, in the case of an absorber at $800$ pc from the GRB.
However, this value should be considered as an upper limit (see text
for details).  }
\label{spe1}
\end{figure}

\begin{table*}[ht]
\caption{\bf Absorption line column densities for four component IV of the main system}
{\footnotesize
\smallskip
\begin{tabular}{|lc|c|}
\hline
Species              & Observed transitions                          & Log N (cm$^{-2}$)   \\
\hline
SiII $^2P^{0}_{1/2}$ &  $\lambda$1526                                & $15.41 \pm 0.22 (S)$\\
\hline
SiII $^2P^{0}_{3/2}$ &  $\lambda$1533                                & $14.25 \pm 0.11   $ \\
\hline
MgI $^1S_0$          &  $\lambda$2026                                & $13.51 \pm 0.08   $ \\
\hline
ZnII $^2S_{1/2}$     &  $\lambda$2026,  $\lambda$2062,               & $12.79 \pm 0.06   $ \\
\hline
CrII $a^6S_{5/2}$    &  $\lambda$2056,  $\lambda$2062, $\lambda$2066 & $13.62 \pm 0.04   $ \\
\hline
NiII $^4F_{9/2}$     &  $\lambda$2166,  $\lambda$2217, $\lambda$2223 & $13.07 \pm 0.04   $ \\
\hline
MnII $a^7S_{3}$      &  $\lambda$2576,  $\lambda$2594, $\lambda$2606 & $13.13 \pm 0.03   $ \\
\hline
FeII $a^6D_{9/2}$    &  $\lambda$2249,  $\lambda$2260, $\lambda$2344,& $14.70 \pm 0.09   $ \\
                     &  $\lambda$2374,  $\lambda$2382, $\lambda$2586,&                     \\
                     &  $\lambda$2600                   &                     \\
\hline
FeII $a^6D_{7/2}$    &  $\lambda$2333,  $\lambda$2365, $\lambda$2389,& $13.76 \pm 0.03   $ \\
                     &  $\lambda$2396,  $\lambda$2599, $\lambda$2612,&                     \\
                     &  $\lambda$2626                   &                     \\
\hline
FeII $a^6D_{5/2}$    &  $\lambda$2328,  $\lambda$2349, $\lambda$2399,& $13.46 \pm 0.04   $ \\
                     &  $\lambda$2405,  $\lambda$2607, $\lambda$2618,&                     \\
                     &  $\lambda$2632                   &                     \\
\hline
FeII $a^6D_{3/2}$    &  $\lambda$2359,  $\lambda$2407, $\lambda$2411,& $13.32 \pm 0.04   $ \\
                     &  $\lambda$2614,  $\lambda$2631                &                     \\
\hline
FeII $a^6D_{1/2}$    &  $\lambda$2411,  $\lambda$2414, $\lambda$2622,& $12.93 \pm 0.05   $ \\
                     &  $\lambda$2629                                &                     \\
\hline
FeII $a^4F_{9/2}$    &  $\lambda$2332,  $\lambda$2348, $\lambda$2360 & $14.04 \pm 0.05   $ \\

\hline
\end{tabular}
}
\end{table*}

\section{The intervening systems}

The absorption lines produced by the gas belonging to the GRB host galaxy
are the dominant features of the high resolution spectrum, but they
are not the only ones. The signal-to-noise level of our spectrum
allowed us to detect lines with equivalent widths as weak as
0.06 \AA, at the $2\sigma$ confidence level. A detailed analysis of
the data found that at least three other absorbers are present along
the line of sight to GRB080330. The two higher redshift ones
($z=1.0179$ and $z=1.0166$) show absorption only from the
{\ion{Mg}{II} $\lambda$2796, $\lambda$2803} doublet, and exhibit a
very simple line profile, that can be fitted by a single Voigt
function.  However, since they are only separated by $\sim 200$ km
s$^{-1}$, they should be interpreted as a single, clumpy system
(fig. 6).  The third absorber is rather complicated.  The absorption
lines detected for this system are the {\ion{Mg}{II} $\lambda$2796,
$\lambda$2803} doublet and the {\ion{Mg}{I} $\lambda$2852} line. This
system has a complex structure and a single Voigt profile is not
adequate to reproduce it. At least five different components are
needed to obtain a good fit to the {\ion{Mg}{II}} data.  The
bluemost component is saturated, and so a reliable column
density for the {\ion{Mg}{II} cannot be obtained. However, this is the
only component that exhibits {\ion{Mg}{I}} absorption at the
90\% confidence level. This {\ion{Mg}{I}} component needs a more
complex modeling to adequately follow its profile. This results in a
further division into six subcomponents (fig. 7). Taking into
account the velocity spread of $\sim260$\,km\,s$^{-1}$, we consider
this system as a single complex absorber.  The redshift path analyzed
for the search of \ion{Mg}{II} systems lies in the intervals $z=0.25 -
0.38$, $z=0.70 - 1.05$, and $z=1.08 - 1.51$.

Table 3 summarizes the rest-frame equivalent widths, column densities,
and redshifts calculated for these intervening systems. Lines 1 and 2
refer to the first double system, while lines 3-7 refer to the more
complex, low redshift one. We report only the total column density for
the bluemost component of {\ion{Mg}{I}} (the corresponding
{\ion{Mg}{II}} is saturated), and do not report the values of all the
six subcomponents. The equivalent width is computed for this system
as a whole.

\begin{table*}[ht]
\caption{\bf Absorption-line rest-frame equivalent widths and column densities for the intervening systems}
{\footnotesize
\smallskip
\begin{tabular}{|lcc|ccc|}
\hline
System&Species         & Transition                    & z       &  EW$\lambda2796$(\AA)  &  N$^a$                       \\
\hline
1&MgII                 &  $\lambda$2796, $\lambda$2803 & 1.0179  & $ 0.06 \pm 0.03$ & $   12.33 \pm 0.07       $  \\
\hline
1&MgII                 &  $\lambda$2796, $\lambda$2803 & 1.0166  & $ 0.08 \pm 0.03$ & $   13.47 \pm 0.19       $  \\
\hline 
2&MgII                 &  $\lambda$2796, $\lambda$2803 & 0.8222  & \multirow{5}{*}{$ 1.70 \pm 0.04$$^b$  }  & $   12.49 \pm 0.03       $  \\
\cline{1-2}
2&MgII                 &  $\lambda$2796, $\lambda$2803 & 0.8220  &                                & $   13.34 \pm 0.03       $  \\
\cline{1-2}
2&MgII                 &  $\lambda$2796, $\lambda$2803 & 0.8219  &                                & $   12.24 \pm 0.04       $  \\
\cline{1-2}
2&MgII                 &  $\lambda$2796, $\lambda$2803 & 0.8216  &                                & $   14.04 \pm 0.09       $  \\
\cline{1-2}
2&MgII                 &  $\lambda$2796, $\lambda$2803 & 0.8213  &                                & $   Sat.                 $  \\
\hline
2&MgI                  &  $\lambda$2852                & 0.8213  & $ 0.79 \pm 0.04$               & $   13.97 \pm 0.15       $  \\

\hline
\end{tabular}

$^a$ Column densities N are logarithmic cm$^{-2}$; Sat marks saturated components.\\
$^b$ The rest equivalenth width of the  $\lambda$2796 transition for the last two components only (z=0.8213 and 0.8216) is $1.30\pm0.04$,. 
}
\end{table*}

\begin{figure}
\centering
\includegraphics[angle=-90,width=9cm]{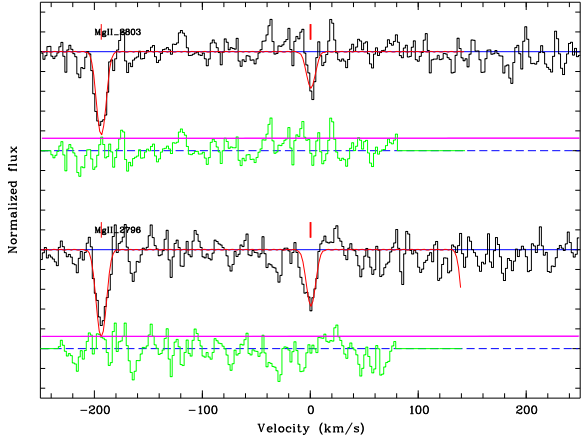}
\caption{The intervening system at z=1.02. {\ion{Mg}{II}} is the
only ion observed in this systems with its $\lambda$2796 and
$\lambda$2803 doublet. Solid lines represent our best-fit
model (see also Table 3). Bottom, solid lines mark the fit residuals.
}
\label{spe1}
\end{figure}

\begin{figure}
\centering
\includegraphics[angle=-90,width=9cm]{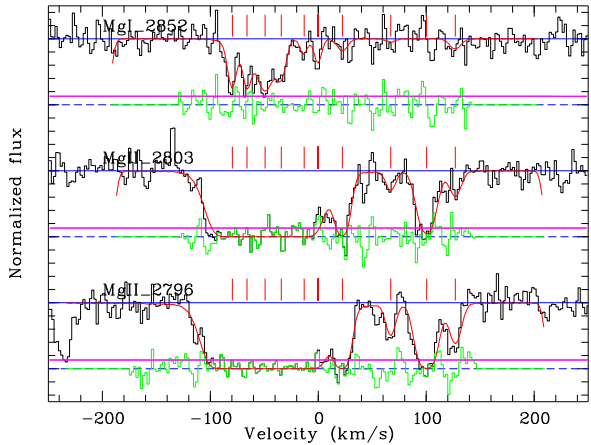}
\caption{The complex intervening system at z=0.8213. A five-component
model was required for the best-fit to the dataset. The neutral
{\ion{Mg}{I}} and the ion {\ion{Mg}{II}} are the only observed
absorbers, with the $\lambda$2852 transition and $\lambda$2796 -
$\lambda$2803 doublet, respectively. {\ion{Mg}{I}} shows significant
absorption in the bluemost component only, which needs a subcomponent
modeling to obtain a good fit. {\ion{Mg}{II}} shows absorption in all
components; the bluemost are saturated (see also Table 3).  Bottom,
solid lines mark the fit residuals.  }
\label{spe1}
\end{figure}

\section{Conclusions}

We have presented high resolution (R=40 000, corresponding to
7.5 km/s) spectroscopy of the optical afterglow of the Gamma Ray Burst
GRB080330, observed using UVES at the VLT $\sim 1.5$ hours after the
trigger.
                                                                                    
We confirm that the redshift of the host galaxy is z=1.51.  The
spectrum shows that the ISM of the GRB host galaxy is complex, with at
least four components contributing to its main absorption system at
z= 1.51. These components are identified in this paper with progressive
numbers of decreasing velocity values.

These absorption lines appear both as neutral metal-absorption lines
({\ion{Mg}{I}}) and low ionization lines ({\ion{Fe}{II}},
{\ion{Si}{II}}, {\ion{Mn}{II}}, {\ion{Cr}{II}}, {\ion{Zn}{II}}).
                                                                                    
In addition, we detect strong {\ion{Si}{II}}, {\ion{Ni}{II}}, and
{\ion{Fe}{II}} excited transitions, but only in component IV (i.e.,
the bluemost component of the main absorption system). Since indirect
UV pumping is likely to be the leading mechanism in exciting these
features (Vreeswijk et al. 2007), we conclude that component IV is the
one more affected by the GRB afterglow, meaning that is the closest to
the GRB progenitor. This distance can be constrained by comparing the
column densities of the ground and excited levels to those predicted
by a time-dependent photoexcitation code for the hypothesis of
indirect UV-pumping. This comparison has been carried out for
{\ion{Fe}{II}} and {\ion{Si}{II}} only; in fact, the column density of
the {\ion{Ni}{II}} ground level could not be measured, because the
corresponding absorption transitions fell in the gap between the blue
and the red regions of the spectrum. We found that the absorbing gas
of component IV is located at $ \sim 280^{+40}_{-50}$ pc from the
GRB. To ensure self-consistency, we ran a photoexcitation code that
confirms that there is a wide range of densities for which the
\ion{Fe}{II} and \ion{Mg}{I} are not substantially photoionized if the
GRB is at $280$ pc. If this is the case, since the abundance ratio
[Cr/Zn] = -0.2 (comparable with DLAs, see Prochaska et al. 2006),
while [Fe/Zn]=-0.9 (among the lowest value for DLAs), the absorber
appears to be very dusty.

Although the distance between the GRB and the absorber is smaller than
the value found from a similar analysis in other GRBs ($1.7$ kpc for
GRB060418, Vreeswijk et al. 2007, and $2-6$ kpc for GRB080319B, D'Elia
et al. 2009), it nevertheless confirms that the power of a GRB affects
a region of gas that is at least a few hundreds of pc in size.

The host galaxy of the GRB is not the only absorber located along the
line of sight to the afterglow. At least two more intervening systems
between z = 1.02 and z = 0.82 have been identified. The system at the
highest redshift can be modeled simply with a double Voigt profile and
exhibits absorption only in the {\ion{Mg}{II} $\lambda$2796,
$\lambda$2803} doublet.  On the other hand, the second is complex,
featuring at least five different components; given the limited
velocity interval however, we assume this to be one single, clumpy
structure along the line of sight. This system exhibits only
{\ion{Mg}{II} $\lambda$2796, $\lambda$2803} doublet and {\ion{Mg}{I}
$\lambda$2852} line absorption.

One of these systems has a rest-frame equivalent width larger than
1\AA. This follows the trend that on average one strong intervening
system is found per afterglow, as has been noted in studies that have
detected  excess of strong {\ion{Mg}{II}} absorbers along GRB
sightlines compared to quasars.

\begin{acknowledgements}
We thank an anonymous referee for several helpful comments.
We acknowledge support from ASI/INAF contracts ASI/I/R/039/04 and
ASI/I/R/023/05/0.
 
\end{acknowledgements}

\end{document}